\documentclass[a4paper,aps,prd,onecolumn,preprintnumbers,showpacs,superscriptaddress,nofootinbib]{revtex4}
\usepackage{graphics,graphicx,epsfig}
\usepackage{amsfonts,amsmath,amssymb}

\begin{document}

\title{Counterterms in semiclassical Ho\v{r}ava-Lifshitz gravity}
\author{Gast\'on Giribet}
\email{gaston@df.uba.ar}
\affiliation{Departamento de F\'{\i}sica, Facultad de Ciencias Exactas y Naturales,
Universidad de Buenos Aires and IFIBA, CONICET. Ciudad Universitaria,
Pabell\'on 1, 1428, Buenos Aires, Argentina.}
\author{Diana  L\'opez Nacir}
\email{dnacir@df.uba.ar}
\affiliation{Departamento de F\'{\i}sica, Facultad de Ciencias Exactas y Naturales,
Universidad de Buenos Aires and IFIBA, CONICET. Ciudad Universitaria,
Pabell\'on 1, 1428, Buenos Aires, Argentina.}
\author{Francisco D. Mazzitelli}
\email{fmazzi@df.uba.ar}
\affiliation{Departamento de F\'{\i}sica, Facultad de Ciencias Exactas y Naturales,
Universidad de Buenos Aires and IFIBA, CONICET. Ciudad Universitaria,
Pabell\'on 1, 1428, Buenos Aires, Argentina.}

\begin{abstract}
We analyze the semiclassical Ho\v{r}ava-Lifshitz gravity for quantum scalar
fields in $3+1$ dimensions. The renormalizability of the theory requires
that the action of the scalar field contains terms with six spatial
derivatives of the field, i.e. in the UV, the classical action of the scalar field
should preserve the anisotropic scaling symmetry ($t \to L^{2z}t,$ $\vec{x}
\to L^2 \vec{x}$, with $z=3$) of the gravitational action. We discuss the
renormalization procedure based on adiabatic subtraction and dimensional
regularization in the weak field approximation. We verify that the divergent
terms in the adiabatic expansion of the expectation value of the
energy-momentum tensor of the scalar field contain up to six spatial
derivatives, but do not contain more than two time derivatives. We compute
explicitly the counterterms needed for the renormalization of the theory up
to second adiabatic order and evaluate the associated $\beta$ functions in
the minimal subtraction scheme.
\end{abstract}

\maketitle


One year and a half ago, Ho\v{r}ava proposed a new approach to formulate a
quantum theory of gravity \cite{Horava}. Ho\v{r}ava's theory, which has
attracted enormous attention, consists of a non-diffeomorphism-invariant
ultraviolet modification of Einstein's general relativity. The main idea in 
\cite{Horava} is to extend Einstein-Hilbert action with higher spatial
derivative terms, whose introduction, while manifestly breaking local
Lorentz invariance, leads to heal the short distance divergences and
ultimately yields a power counting renormalizable theory. The way this is
achieved without introducing ghost instabilities is keeping the requirement
of the theory to be of second-order in time derivatives. This introduces an
asymmetry between the time coordinate $t$ and the coordinates $x^{i}$
associated to a preferable foliation that defines a three-dimensional
space-like hypersurface of induced metric $^{(3)}g_{ij}$. In turn,
four-dimensional diffeomorphism invariance results manifestly broken at
short distances, and consequently the theory only exhibits diffeomorphism
invariance in three-dimensions, in addition to the reparameterization
invariance in time. According to this picture, the four-dimensional general
covariance of gravity would emerge merely as an approximate symmetry at low
energy.

Fragmentation of space-time diffeomorphism invariance in the form of a
preferable three-dimensional space-like hypersurfaces defined at constant
time, immediately suggests to consider the ADM decomposition for the metric
as the convenient picture. Namely, consider 
\begin{equation}
ds^{2}=-N^{2}dt^{2}+g_{ij}(dx^{i}+N^{i}dt)(dx^{j}+N^{j}dt),  \label{ADM}
\end{equation}%
where, as usual, Latin indices refer to the spatial coordinates, $i,j=1,2,3$%
, and $g_{ij}=$ $^{(3)}g_{ij}$. In the non-projectable theory, the lapse
function $N$ depends both on time and the spatial coordinates, in such a way
general relativiy is captured within this formulation.

The action of Ho\v{r}ava's theory is given by 
\begin{equation}
S=\frac{1}{16\pi G}\int dt\text{ }dx^{3}\text{ }N\sqrt{g}\left(
K_{ij}K^{ij}-\lambda K^{2}-2\Lambda+\xi R-V\right)  \label{S}
\end{equation}%
where $\Lambda$ is the bare cosmological constant and $\lambda $ and $\xi $
are arbitrary bare coupling constants; Einstein theory corresponds to the
special choice $\lambda =1.$ The extrinsic curvature $K_{ij}$ in the ADM
variables takes the form $K_{ij}=(\overset{.}{g}_{ij}+\nabla
_{i}N_{j}+\nabla _{j}N_{i})/(2N)$, whose trace is given by $K=K_{ij}g^{ij}$.
Here, $\nabla _i$ denotes the spatial derivative, while the dot denotes the
derivative with respect to time. The function $V$ in (\ref{S}) plays the r%
\^{o}le of a potential, as it only depends on spatial derivatives of the
metric, which would include higher derivative contributions (see below). In (%
\ref{S}), $R$ represents the Ricci scalar curvature of the three-dimensional
space-like hypersurface of induced metric $g_{ij}$.

The presence of terms in the action that involve higher spatial derivatives
leads to different scaling dimensions for the time and the spatial
coordinates. This is represented by the scaling symmetry

\begin{equation}
x^{i}\rightarrow L^{2}\text{ }x^{i},\qquad t\rightarrow L^{2z}\text{ }%
t,\qquad N\rightarrow N,\qquad N_{i}\rightarrow L^{-4z}N_{i}  \label{scaling}
\end{equation}%
which is characterized by the dynamical critical exponent $z$. A consistent
choice is $z=3$, which is the one we will consider throughout this paper.
With this choice, we can consider the potential%
\begin{equation}
V=\frac{1}{2}(a_{1}\Delta R+a_{2}R_{ij}R^{ij}+...\ )+4\pi G(b_{1}\Delta
^{2}R+b_{2}R_{ij}R^{jk}R_{k}^{i}+...\ ),  \label{V}
\end{equation}%
where we are using the notation $\Delta =\nabla _{i}\nabla ^{i}$. The
ellipses in (\ref{V}) stand for other terms of the same dimension.

At low energy, the action turns out to be dominated by the term that
involves the Ricci scalar $R$, with coefficient $\xi $. In turn, the theory
would reproduce Einstein's general relativity in the infrared, provided $%
\lambda $ flows to the value $\lambda _{\text{IR}}=1$. The consistency of
the theory and the validity of this hypothesis were extensively discussed in
the literature; see \cite{Charmousis, Li, Blas,Henneaux} and references therein. Of
special interest is the discussion in \cite{Blas2, Blas3}, where an improved
version of Ho\v{r}ava gravity, which seems to be free of pathologies, was
presented.

About renormalizability, of particular importance is the question on how the
coupling of Ho\v{r}ava gravity to matter affects the properties of the
theory in the UV. With the purpose of addressing this problem, we study the
coupling of the theory to a quantum scalar field, representing the matter
content. The gravitational field will be treated at a classical level, so we
are considering a semiclassical Ho\v{r}ava-Lifshitz gravity. It is
interesting to remark that, if the matter fields satisfy the usual
dispersion relations (i.e if the classical action has four-dimensional
general covariance), the theory is non renormalizable. Indeed, it is well
known in the context of quantum field theory in curved spacetimes that in
order to absorb the divergences associated to the matter fields it is
necessary to include in the gravitational action terms proportional to $%
\mathcal{R}^2, \mathcal{R}_{\mu\nu}\mathcal{R}^{\mu\nu}$ and $\mathcal{R}%
_{\mu\nu\rho\sigma}\mathcal{R}^{\mu\nu\rho\sigma}$, where $\mathcal{R}%
_{\mu\nu\rho\sigma}$ denotes the components of the four-dimensional Riemann
tensor. These terms contain  four time derivatives of the metric, and
therefore are not included in Ho\v{r}ava gravity. As we will see,
renormalizability of the field theory demands that the action for the matter
sector contains terms with six spatial derivatives, implying that in the UV the
coupling to the scalar field preserves the Lifshitz-type anisotropic scaling
with critical exponent $z=3$. We will verify that the divergent terms in the
adiabatic expansion of the expectation value of the stress-tensor associated
to the scalar field actually contains up to six spatial derivatives but it
remains of second order in time derivatives. We will explicitly compute the
counterterms needed for the renormalization of the theory up to second
adiabatic order, and we will write down the corresponding $\beta $-functions
in the minimal substraction scheme.

The computation techniques we will employ here have recently been employed
with success to study renormalization in the so-called Einstein-aether
theory and in other field theories with modified dispersion relations \cite%
{DiegoDiana}. The idea for using the same
techniques in Ho\v{r}ava-Lifshitz gravity comes from the observation that
this theory is closely related to such Lorentz violating scenarios; see for
instance \cite{Jacobson} and \cite{Germani}.

Let us begin by considering the coupling of Ho\v{r}ava-Lifshitz gravity to a
Lifshitz-type scalar field. In the ADM form, the components $^{(4)}g_{\mu
\nu }$ of the four-dimensional metric (\ref{ADM}) are given by%
\begin{equation*}
^{(4)}g_{00}=-N^{2}+g_{ij}N^{i}N^{j},\qquad ^{(4)}g_{0i}=g_{ij}N^{j},\qquad
^{(4)}g_{ij}=g_{ij},
\end{equation*}%
where $i,j=1,2,3$, and $g_{ij}$ refers to the metric on the
three-dimensional foliation of constant $t$. We consider small perturbations
of the metric about flat space; namely, we write 
\begin{equation}
N=1+\delta n,\qquad N^{i}=\delta N^{i},\qquad g_{ij}=\delta _{ij}+h_{ij}.
\label{h}
\end{equation}

We consider a matter Lagrangian giving by a scalar field $\varphi $ that
also exhibits anisotropic critical scaling; namely\cite{Brandenberger}%
\begin{equation*}
S_{\varphi }=\int dtdx^{3}\sqrt{g}N\left( \frac{1}{2N^{2}}(\overset{.}{%
\varphi }-N^{i}\partial _{i}\varphi )^{2}+F(\varphi ,\partial \varphi )-%
\frac{1}{2}m^{2}\varphi ^{2}\right)
\end{equation*}%
where the potential $F(\varphi ,\partial \varphi )$ is given by%
\begin{equation*}
F(\varphi ,\partial \varphi )=-g_{1}\partial ^{i}\varphi \partial
_{i}\varphi -g_{2}\left( \Delta \varphi \right) ^{2}+g_{3}\Delta ^{2}\varphi
\Delta \varphi ,
\end{equation*}%
where we have to be reminded of the definition $\Delta \varphi =\frac{1}{%
\sqrt{g}}\partial _{i}(\sqrt{g}g^{ij}\partial _{j}\varphi )$, with $g^{ij}=$ 
$^{(3)}g^{ij}=$ $^{(4)}g^{ij}-$ $(^{(4)}g^{i0}$ $^{(4)}g^{j0})/^{(4)}g^{00}$%
, and $^{(4)}g^{00}=-N^{-2}$. The equation for the Green function reads%
\begin{eqnarray}
&& -\partial _{t}\left( \frac{\sqrt{g}}{N}(\partial _{t}-N^{i}\text{ }%
\partial _{i})G(x,x^{\prime })\right) +\partial _{j}\left( \frac{N^{i}\sqrt{g%
}}{N}(\partial _{t}-N^{i}\text{ }\partial _{i})G(x,x^{\prime })\right) -m^{2}%
\sqrt{g}N\text{ }G(x,x^{\prime })+ 2g_{1}\partial _{i}\left( N\sqrt{g}\text{ 
}\partial ^{i}G(x,x^{\prime })\right) -  \notag \\
&& 2g_{2}\sqrt{g}\Delta (N\text{ }\Delta G(x,x^{\prime })) + g_{3} \sqrt{g}%
\Delta \left( N\text{ }\Delta ^{2}G(x,x^{\prime })\right) +g_{3}\sqrt{g}%
\Delta ^{2}\left( N\text{ }\Delta G(x,x^{\prime })\right) =-\delta
(x-x^{\prime }).  \label{G}
\end{eqnarray}

At the linearized level we have (\ref{h}), which yields $\sqrt{g}=1+h/2$, $%
g^{ij}=\delta _{ij}-h_{ij}$. This can be used to write the equation for the
Green function (\ref{G}) in the weak field approximation. The Feynman
propagator of zero order in the metric perturbations reads%
\begin{equation*}
G_{F}^{(0)}(x,x^{\prime })=\int \frac{d^{4}k}{(2\pi )^{4}}\frac{e^{ik\cdot
(x-x^{\prime })}}{\left( -k_{0}^{2}+\omega _{k}^{2}-i\varepsilon \right) },
\end{equation*}%
where $k=|\vec{k}|$ and%
\begin{equation*}
\omega _{k}^{2}=m^{2}+2g_{1}k^{2}+2g_{2}k^{4}+2g_{3}k^{6},
\end{equation*}%
while the first order contribution can be written as follows%
\begin{eqnarray}
G_{F}^{(1)}(x,x^{\prime }) &=&\int \frac{d^{4}k}{(2\pi )^{4}}\int \frac{%
d^{4}p}{(2\pi )^{4}}\frac{e^{ip\cdot x}e^{ik\cdot (x-x^{\prime })}\text{ }%
f_{k}(p)}{\left( -k_{0}^{2}+\omega _{k}^{2}-i\varepsilon \right) \left(
-(k_{0}+p_{0})^{2}+\omega _{|\vec{k}+\vec{p}|}^{2}\right) }  \notag
\label{Green1} \\
&\equiv &\int \frac{d^{4}k}{(2\pi )^{4}}\int \frac{d^{4}p}{(2\pi )^{4}}\frac{%
e^{ip\cdot x}e^{ik\cdot (x-x^{\prime })}\text{ }f_{k}(p)\left( 1+\epsilon
_{p}\right) ^{-1}}{\left( -k_{0}^{2}+\omega _{k}^{2}-i\varepsilon \right)
\left( -k_{0}^{2}+\omega _{k}^{2}\right) }.
\end{eqnarray}%
Here, $f_{k}(p)$ is a function of $k_{0},k_{i},p_{0}$, and $p_{i}$ that is
linear in the metric perturbations 
\begin{eqnarray}
f_{k}(p) &=&\left( \delta n-\frac{h}{2}\right) k_{4}^{2}+2ik_{4}k_{i}\delta
N^{i}-\left( \delta n+\frac{h}{2}\right) \omega _{k}^{2}+h_{ij}k^{i}k^{j}%
\frac{d\omega _{k}^{2}}{dk^{2}}+h_{ij}\delta _{rs}k^{i}k^{j}k^{r}p^{s}\frac{%
d^{2}\omega _{k}^{2}}{d(k^{2})^{2}}  \notag \\
&&+ik_{4}p_{0}\left( \delta n-\frac{h}{2}\right) -\delta
N^{i}p_{0}k_{i}+ik_{4}\delta N^{i}p_{i}-\left( \left( \delta n+\frac{h}{2}%
\right) \delta _{ij}\text{ }p^{i}k^{j}+h^{ij}p_{i}k_{j}\right) \frac{d\omega
_{k}^{2}}{dk^{2}}  \notag \\
&&-\frac{d^{2}\omega _{k}^{2}}{d(k^{2})^{2}}\left( \frac{\delta n}{2}%
p^{2}k^{2}+\frac{h}{2}\left( \delta _{ij}k^{i}p^{j}\right) ^{2}-\frac{1}{2}%
h_{ij}k^{i}k^{j}p^{2}-h_{ij}\delta _{rs}p^{i}p^{r}k^{j}k^{s}\right)   \notag
\\
&&+\frac{d^{3}\omega _{k}^{2}}{d(k^{2})^{3}}\left( \frac{\delta n}{4}%
p^{2}k^{4}+\frac{2}{3}h_{ij}k^{i}k^{j}(\delta _{rs}k^{r}p^{s})^{2}-\frac{%
\delta n}{3}(\delta _{ij}k^{i}p^{j})^{2}k^{2}\right) -\frac{d^{2}\omega
_{k}^{2}}{d(k^{2})^{2}}\left( \frac{h}{4}p^{2}\delta _{ij}k^{i}p^{j}-\frac{1%
}{2}h_{ij}k^{i}p^{j}\right)   \notag \\
&&-\frac{d^{3}\omega _{k}^{2}}{d(k^{2})^{3}}\left( \frac{\delta n}{3}%
p^{2}k^{2}\delta _{ij}k^{i}p^{j}+\frac{h}{3}\left( \delta
_{ij}k^{i}p^{j}\right) ^{3}-\frac{2}{3}h_{ij}k^{i}k^{j}p^{2}\delta
_{rs}k^{r}p^{s}-\frac{2}{3}h_{ij}p^{i}k^{j}(\delta
_{rs}k^{r}p^{s})^{2}\right)   \notag \\
&&-\frac{d^{3}\omega _{k}^{2}}{d(k^{2})^{3}}\left( \frac{\delta n}{12}%
p^{4}k^{2}+\frac{h}{3}p^{2}\left( \delta _{ij}k^{i}p^{j}\right) ^{2}-\frac{1%
}{6}h_{ij}k^{i}k^{j}p^{4}-\frac{2}{3}h_{ij}\delta
_{rs}p^{i}p^{r}k^{j}k^{s}p^{2}\right)   \notag \\
&&-\frac{d^{3}\omega ^{2}}{d(k^{2})^{3}}\left( \frac{h}{12}p^{4}\delta
_{ij}k^{i}p^{j}-\frac{1}{6}h_{ij}k^{i}p^{j}p^{4}\right) .  \label{fkp}
\end{eqnarray}%
with  $p^{2}=|\vec{p}|^{2}$, and $\epsilon _{p}$ is defined as 
\begin{equation}
\epsilon _{p}=\frac{-2k_{0}p_{0}-p_{0}^{2}+\omega _{|\vec{k}+\vec{p}%
|}^{2}-\omega _{k}^{2}}{-k_{0}^{2}+\omega _{k}^{2}}.  \label{eps}
\end{equation}%
In what follows, for the sake of convenience, we perform a Wick rotation in
Eq.(\ref{Green1}) and we call $k_{4}=ik_{0}$.

To obtain the adiabatic expansion of the Feynman propagator we start by
expanding the integrand of Eq.(\ref{Green1}) in powers of $p_{0}$ and $p_{i}$%
. The different adiabatic orders of $f_{k}(p)$ are given by 
\begin{subequations}
\label{fk}
\begin{align}
f_{k}^{\text{ad}(0)}=& \left( \delta n-\frac{h}{2}\right)
k_{4}^{2}+2ik_{4}k_{i}\delta N^{i}-\left( \delta n+\frac{h}{2}\right) \omega
_{k}^{2}+h_{ij}k^{i}k^{j}\frac{d\omega _{k}^{2}}{dk^{2}}, \\
f_{k}^{\text{ad}(1)}=& ik_{4}p_{0}\left( \delta n-\frac{h}{2}\right) -\delta
N^{i}p_{0}k_{i}+ik_{4}\delta N^{i}p_{i}-\left( \left( \delta n+\frac{h}{2}%
\right) \delta _{ij}\text{ }p^{i}k^{j}+h^{ij}p_{i}k_{j}\right) \frac{d\omega
_{k}^{2}}{dk^{2}}  \notag \\
+& h_{ij}\delta _{rs}k^{i}k^{j}k^{r}p^{s}\frac{d^{2}\omega _{k}^{2}}{%
d(k^{2})^{2}}, \\
f_{k}^{\text{ad}(2)}=& -\frac{d^{2}\omega _{k}^{2}}{d(k^{2})^{2}}\left( 
\frac{\delta n}{2}p^{2}k^{2}+\frac{h}{2}\left( \delta _{ij}k^{i}p^{j}\right)
^{2}-\frac{1}{2}h_{ij}k^{i}k^{j}p^{2}-h_{ij}\delta
_{rs}p^{i}p^{r}k^{j}k^{s}\right)  \notag \\
+& \frac{d^{3}\omega _{k}^{2}}{d(k^{2})^{3}}\left( \frac{\delta n}{4}%
p^{2}k^{4}+\frac{2}{3}h_{ij}k^{i}k^{j}(\delta _{rs}k^{r}p^{s})^{2}-\frac{%
\delta n}{3}(\delta _{ij}k^{i}p^{j})^{2}k^{2}\right) .
\end{align}%
From Eq.(\ref{fkp}) one can see that no powers of $p_{0}$ appear in the
adiabatic orders $f_{k}^{\text{ad}(m)}$ with $m\geq 3$. Besides this
property, these $m\geq 3$ adiabatic orders are not relevant for our
discussion. Note also that the adiabatic orders $f_{k}^{\text{ad}(m)}$
with $m\geq 6$ vanish.

We have also to expand $\epsilon _{p}$ in its adiabatic orders. That is, 
\end{subequations}
\begin{subequations}
\label{epsilonp}
\begin{align}
\epsilon _{p}^{\text{ad}(1)}=& \frac{2ik_{4}p_{0}+2\frac{d\omega _{k}^{2}}{%
dk^{2}}\text{ }\delta _{ij}k^{i}p^{j}}{\omega _{k}^{2}+k_{4}^{2}}, \\
\epsilon _{p}^{\text{ad}(2)}=& \frac{-p_{0}^{2}+\frac{d\omega _{k}^{2}}{%
dk^{2}}\text{ }p^{2}+2\frac{d^{2}\omega _{k}^{2}}{d(k^{2})^{2}}\text{ }%
(\delta _{ij}k^{i}p^{j})^{2}}{\omega _{k}^{2}+k_{4}^{2}},
\end{align}%
%
%
and $\epsilon _{p}^{\text{ad}(0)}=0$. It is easy to see that $\epsilon _{p}^{%
\text{ad}(m)}$ with $m\geq 3$ do not involve powers of $p_{0}$; explicit
expressions of these adiabatic orders are not necessary for our present
purposes.

The expression from which one obtains the components $\langle  T_{\mu \nu }(x)\rangle $ of
the expectation value of stress tensor after taking the coincidence limit $x\rightarrow x^{\prime
}$ is obtained by evaluating the derivatives of the propagator $%
G_{F}(x,x^{\prime })$, as usual. Appropriate regularization turns out to be
necessary. For the case of the component $T_{00}(x)$, we find 
\end{subequations}
\begin{eqnarray}
\langle T_{00}(x)\rangle &=&\lim_{x\rightarrow x^{\prime }}\left\{ \partial
_{t}\partial _{t^{\prime }}+\frac{1}{2}(\delta N^{i}\partial _{i}\partial
_{t^{\prime }}+\delta N^{i^{\prime }}\partial _{i^{\prime }}\partial _{t})+%
\frac{m^{2}}{2}(1+2\delta n)\text{ }\right.  \notag \\
&-&(1+2\delta n)\left( -g_{1}\delta ^{ii^{\prime }}\partial _{i}\partial
_{i}^{\prime }-g_{2}\partial ^{2}\partial ^{\prime 2}+\frac{1}{2}%
g_{3}(\partial ^{4}\partial ^{\prime 2}+\partial ^{2}\partial ^{\prime
4})\right) \text{ }  \notag \\
&+&\left( g_{1}(h^{ij^{\prime }}\partial _{i}\partial _{j^{\prime
}}+h^{i^{\prime }j}\partial _{i^{\prime }}\partial _{j})+g_{2}\left(
h^{ij}\partial _{i}\partial _{j}\partial ^{\prime 2}+h^{i^{\prime }j^{\prime
}}\partial _{i^{\prime }}\partial _{j^{\prime }}\partial ^{2}\right)
-g_{2}(\partial _{i}\overline{h}_{ij}\partial _{j}\partial ^{\prime
2}+\partial _{i}\overline{h}_{ij^{\prime }}\partial ^{2}\partial _{j^{\prime
}})\right) \text{ }  \notag \\
&+&\frac{1}{2}g_{3}\left( h_{ij}\partial _{i}\partial _{j}\partial ^{\prime
4}+h_{i^{\prime }j^{\prime }}\partial _{i^{\prime }}\partial _{j^{\prime
}}\partial ^{4}+\partial _{i}\overline{h}_{ij}\partial _{j}\partial ^{\prime
4}+\partial _{i}\overline{h}_{ij^{\prime }}\partial _{j^{\prime }}\partial
^{4}\right) \text{ }  \notag \\
&-&g_{3}\left( h^{i^{\prime }j^{\prime }}\partial ^{2}\partial _{i^{\prime
}}\partial _{j^{\prime }}\partial ^{\prime 2}+h^{ij}\partial ^{\prime
2}\partial _{i}\partial _{j}\partial ^{2}-\partial _{i}\overline{h}%
_{ij}\partial _{j}\partial ^{2}\partial ^{\prime 2}-\partial _{i}\overline{h}%
_{ij^{\prime }}\partial ^{2}\partial _{j^{\prime }}\partial ^{\prime
2}\right) \text{ }  \notag \\
&-&\frac{1}{2}g_{3}\left( \partial ^{2}h^{ij}\partial _{i}\partial
_{j}\partial ^{\prime 2}+\partial ^{2}h^{i^{\prime }j^{\prime }}\partial
^{2}\partial _{i^{\prime }}\partial _{j^{\prime }}-\partial _{i}\partial ^{2}%
\overline{h}_{ij}\partial _{j}\partial ^{\prime 2}-\partial _{i}\partial ^{2}%
\overline{h}_{ij^{\prime }}\partial ^{2}\partial _{j^{\prime }}\right) \text{
}  \notag \\
&-&\left. g_{3}\left( \partial _{k}h^{ij}\partial _{i}\partial _{j}\partial
_{k}\partial ^{\prime 2}+\partial _{k^{\prime }}h^{i^{\prime }j^{\prime
}}\partial ^{2}\partial _{i^{\prime }}\partial _{j^{\prime }}\partial
_{k^{\prime }}-\partial _{i}\partial _{k}\overline{h}_{ij}\partial
_{k}\partial _{j}\partial ^{\prime 2}-\partial _{i}\partial _{k^{\prime }}%
\overline{h}_{ij^{\prime }}\partial ^{2}\partial _{k^{\prime }}\partial
_{j^{\prime }}\right) \text{ }\right\} \text{Im }G_{F}(x,x^{\prime })
\label{t00}
\end{eqnarray}%
where $\overline{h}_{ij}=h_{ij}-\frac{h}{2}\delta _{ij}$,  $\partial^2=\partial_i\partial_i$,  and  a primed
index on a derivative indicates that the derivative is taken with respect to
a primed coordinate.


For the sake of simplicity, and because it is enough for our present
purposes, we partially fix the gauge by setting $\delta N^{i}=0$. This
greatly simplifies the expression of the $\langle T_{0i}(x) \rangle $ component, which reads%
\begin{equation}  \label{t0i}
\langle T_{0i}(x)\rangle =\frac{1}{2} \lim_{x\to
x^{\prime}}(\partial_t\partial_{i^{\prime}}+\partial_{t^{\prime}}%
\partial_i) \text{ Im }G_{F}(x,x^{\prime }).
\end{equation}

To obtain the regularized expectation values of the stress tensor we use
dimensional regularization. Therefore, after computing the derivatives of $%
\text{ Im }G_{F}(x,x^{\prime })$ that appear in Eqns.(\ref{t00}) and (\ref%
{t0i}), we can set $x=x^{\prime }$. Then, it is straightforward to separate
the different adiabatic orders of $\langle T_{0\mu }(x)\rangle $ $(\mu
=0,1,2,3) $, before performing the integrations. In this way we obtain an
integral expression for each adiabatic order. The next step is to use
dimensional regularization to perform the integrals. We apply dimensional
regularization to both temporal and spatial directions, but in a separated
way. That is, we split the $d-$dimensional integrals into integrals in $d_{1}
$ and $d_{2}$ dimensions, with $d=d_{1}+d_{2}$, where $d_{1}\rightarrow 1$
and $d_{2}\rightarrow 3$. All integrals in $k_{4}$ are of the form 
\begin{equation*}
I_{d_{1}}(j,l)=\frac{\Omega _{d_{1}}}{(2\pi )^{d_{1}}}\int_{0}^{+\infty
}dk_{4}\frac{k_{4}^{d_{1}-1+j}}{(k_{4}^{2}+\omega _{k}^{2})^{l}}=\frac{%
\Omega _{d_{1}}}{(2\pi )^{d_{1}}}\frac{\omega _{k}^{-2l+j+d_{1}}}{2\Gamma (l)%
}\Gamma \left( l-\frac{j+d_{1}}{2}\right) \Gamma \left( \frac{j+d_{1}}{2}%
\right)
\end{equation*}%
where $j$ is an even number ($I_{d_{1}}(j,l)=0$ if $j$ is odd) and $l$ is an
integer; $\Omega _{d_{1}}=2\pi ^{d_{1}/2}/\Gamma (d_{1}/2)$, with $\Gamma
(z) $ being the Gamma function. Note that the right hand side is finite in
the limit $d_{1}\rightarrow 1$ (as is usual in dimensional regularization
for odd dimensions), then, we set $d_{1}=1$.

Using the adiabatic expansion of the expectation value of the stress-tensor,
by simple power counting one can study up to which adiabatic order it
contains divergences and how many temporal derivatives do appear in the
divergent terms. To illustrate this, let us consider as an example the
following contribution to $\langle T_{00}\rangle $: 
\begin{equation*}
t_{00}(x)=\lim_{x\rightarrow x^{\prime }}\partial _{t}\partial _{t^{\prime }}%
\text{Im }G_{F}(x,x^{\prime }).
\end{equation*}%
It can be shown that this term, along with others which are similar, are the
most divergent ones. The contribution that is linear in the metric
perturbations can be written as 
\begin{equation}
t_{00}(x)=-\int \frac{d^{4}p}{(2\pi )^{4}}e^{ip\cdot x}\int \frac{d^{d}k_{E}%
}{(2\pi )^{d}}\frac{k_{4}(k_{4}+ip_{0})}{(k_{4}^{2}+\omega _{k}^{2})^{2}}%
f_{k}(p)\sum_{r=0}(-\epsilon _{p})^{r},  \label{t00ex}
\end{equation}

Let us first analyze the terms that do not contain $p_0$. In a schematic
way, it is simple to show that the ultraviolet behavior of a term in $%
f_{k}(p)$ and in $\epsilon _{p}$, respectively, is given by $\epsilon
_{p}\sim k^{6-n}p^{n}/(k_{4}^{2}+\omega _{k}^{2})$, with $1\leq n\leq 6$,
and $f_{k}(p)\sim \delta g\,k^{6-s}p^{s}$, where $0\leq s\leq 5$ and $\delta
g$ represents a component of $\delta ^{(4)}g_{\mu \nu }$. Then, for a term
characterized by $n$, $s$ and $r$, the integration in $k_{4}$ yields 
\begin{equation*}
\int {d^{d_{1}}k_{E}}\frac{k_{4}^{2}f_{k}(p)}{(k_{4}^{2}+\omega _{k}^{2})^{2}%
}(-\epsilon _{p})^{r}\sim \delta g\int d^{d_{1}}k_{E}\frac{%
k_{4}^{2}k^{6(1+r)}}{(k_{4}^{2}+\omega _{k}^{2})^{2+r}}\left( \frac{p}{k}%
\right) ^{s+nr}\sim \delta g\,k^{3}\left( \frac{p}{k}\right) ^{s+nr}.
\end{equation*}%
Therefore, by power counting it can be shown that the integral in $k$ is
convergent only if $s+rn>d+2$. That is, for $d=4$, we have that $%
p^{d+2}=p^{6}$ is the maximum power of $p$ that appears in a divergent
contribution; i. e., terms of adiabatic order greater than six are finite.

Let us now analyze the terms that contain powers of $p_0$. 
In Eq.(\ref{t00ex}) $p_0$ appears explicitly and also implicitly through $%
f_k(p)$ and $\epsilon_p$. Notice that of all the adiabatic orders only $%
f_k^{ad(1)}(p)$, $\epsilon_p^{ad(1)}$ and $\epsilon_p^{ad(2)}$ depend on $%
p_0 $ (see Eqns.(\ref{fk}) and (\ref{epsilonp}) and the paragraphs that
follows each of them). Then, one can write all the terms of second adiabatic
order that involve one or two powers of $p_0$ (i.e., terms with $p_0 p_i$ or
with $p_0^2$ and no additional power of $p_{\mu}$) and show that all of them
are logarithmically divergent. Hence, as the convergence improves with the
adiabatic order, we can conclude that the contribution of higher adiabatic
orders with at least one power of $p_0$ will be finite. Below, we compute
explicitly the terms of second adiabatic order of the $00$ and $0i$
components of $\langle T_{\mu\nu}\rangle$ and we show that while the terms
with one and two powers of $p_0$ are both logarithmically divergent, the
ones with $p_0^2$ do not appear in the final result. We expect that terms
with $p_0^2$ do appear in the $ij$ components, but here we will not compute
these explicitly.

In summary, in $\langle
T_{\mu\nu} \rangle$  there appear divergences up to in the sixth  {\it weighted adiabatic order}, where the weighted adiabatic order of a term is given by  ($z=3$)
$$W=z n_0+n_i$$ 
  where $n_i$ and $n_0$  are, respectively,  the number of spatial derivatives and time derivatives appearing in the  term.  This is analogous to the weighted power counting criterion introduced in  \cite{Anselmi} for  field theories in Minkowski spacetime (see also \cite{Visser}).

In order to illustrate the procedure by which we obtain the regularized
adiabatic orders in terms of $3-$tensorial quantities, let us consider as an
example the zeroth adiabatic order of $\langle T_{00}(x)\rangle$. After
performing the integrals in $p_{\mu}$ (which are straightforward) and the
integral in $k_4$ (as described above), we obtain 
\begin{equation}
\langle T_{00}(x)\rangle^{\text{ad}(0)} =\frac{\mu^{4-d}}{2} \int \frac{%
d^{d_2}k}{(2\pi)^{d_2}}\left\{ \omega_{k}\left(1+2\delta n-\frac{h}{2}
\right)-\frac{h_{ij}}{\omega_{k}}\left(g_1 k_i k_j+2 g_2 k_i k_j k^2+3 g_3
k_i k_j k^4\right)\right\},
\end{equation}
where $\mu$ is an arbitrary parameter of dimensions of mass, introduced to
ensure that $\varphi$ has the correct dimensionality.

To carry out the angular integrations we use the following property \cite%
{CollinsRen}: 
\begin{equation*}
\int d^{d_2}k k^{i_1} ... k^{i_r} g(k^2)= 
\begin{cases}
0 & \text{if $r$ is odd,} \\ 
T^{i_1...i_{r}} A_{r}[g] & \text{if $r$ is even,}%
\end{cases}%
\end{equation*}
where 
\begin{subequations}
\begin{align}
T^{i_1...i_{r}}&=\frac{1}{r!}[\delta^{i_1 i_2}\delta^{i_3
i_4}...\delta^{i_{r-1} i_r}+\hbox{all permutations of the $i$'s} ],  \notag
\\
A_{r}[g]&=\frac{2\pi^{d_2/2}\Gamma[(r+1)/2]}{\Gamma[1/2]\Gamma[(d_2+r)/2]}%
\int_0^{\infty} dk k^{d_2+r-1} g(k^2).  \notag
\end{align}%
The remaining integrals can be related by performing an integration by
parts, 
\end{subequations}
\begin{eqnarray}
\langle T_{00}(x)\rangle^{\text{ad}(0)}& =&\frac{\mu^{4-d}\Omega_{d_2}}{4
(2\pi)^{d_2}} \int_0^{+\infty} dk^2 k^{d_2-2} \left\{
\omega_{k}\left(1+2\delta n-\frac{h}{2} \right)-\frac{h}{d_2}k^2 \frac{%
d\omega_{k}}{dk^2}\right\}  \notag \\
&=&\frac{\mu^{4-d}\Omega_{d_2}}{4 (2\pi)^{d_2}} (1+2\delta
n)\int_0^{+\infty} dk^2 k^{d_2-2} \omega_{k}={-}^{(4)}g_{00} \frac{%
\mu^{4-d}\Omega_{d_2}}{4 (2\pi)^{d_2}}I_0.  \label{t00ad0}
\end{eqnarray}
where we have defined $I_0=\int_0^{+\infty} dk^2 k^{d_2-2} \omega_{k}$.
Moreover, one can easily show that, due to the gauge condition $\delta N^i=0$%
, one has $\langle T_{0i}(x)\rangle^{\text{ad}(0)}=0$. Therefore, as
expected, the lowest adiabatic order of the energy momentum tensor is
proportional to the metric and can be absorbed into a redefinition of the
cosmological constant (see below).

We follow the same procedure for the second adiabatic order of $\langle
T_{0\mu}(x)\rangle$. After a long but straightforward calculation the
results are 
\begin{eqnarray}  \label{T0iad2}
\langle T_{00}(x)\rangle^{\text{ad}(2)}&=& -\frac{\mu^{4-d}\Omega_{d_2}}{48
(2\pi)^{d_2}}I_1 (\partial_i\partial_j h_{ij}-\partial^2h)=-\frac{%
\mu^{4-d}\Omega_{d_2}}{24 (2\pi)^{d_2}}I_1 G_{00},  \notag \\
\langle T_{0i}(x)\rangle^{\text{ad}(2)}&=& -\frac{\mu^{4-d}\Omega_{d_2}}{48
(2\pi)^{d_2}} \left\{I_2 (\partial_j\dot{h}_{ij}-\partial_i\dot{h})+\frac{I_3%
}{d_2 (d_2+2)}(2\partial_j \dot{h}_{ij}+\partial_i\dot{h})\right\}  \notag \\
&=&-\frac{\mu^{4-d}\Omega_{d_2}}{24 (2\pi)^{d_2}} \left\{I_2 G_{0i}+\frac{I_3%
}{d_2 (d_2+2)}(2 G_{0i}+3\partial_iK^j_j)\right\},
\end{eqnarray}
where $G_{00}$ and $G_{0i}$ are components of the linearized Einstein tensor 
$G_{\mu\nu}$ and $K^i_j$ is the linearized extrinsic curvature, and we have
defined the following integrals: 
\begin{equation*}
I_1=\int_0^{+\infty}dk^2 \frac{k^{d_2-2}}{\omega_k}\frac{d\omega_k^2}{dk^2}%
,\,\, I_2=\int_0^{+\infty}dk^2 \frac{k^{d_2-2}}{\omega_k},\,\,
I_3=\int_0^{+\infty}dk^2 \frac{k^{d_2+2}}{\omega_k^3}\frac{d^2\omega_k^2}{%
d(k^2)^2}.
\end{equation*}
Notice that while $I_1$ is quartically divergent, $I_2$ and $I_3$ are
logarithmically divergent. The terms in Eq.(\ref{T0iad2}) have first time
derivatives of the metric and, as we have anticipated, result to be
logarithmically divergent. We have repeated all the calculations without
partially fixing the gauge $\delta N^i=0$, and reobtained Eq. (\ref{T0iad2})
as a cross-check.

With these results we can now analyze the renormalization of the bare
constants associated to the terms of second adiabatic order that appear in
the gravitational action (\ref{S}). To do so, we start by writing the $00$
and $0i$ parts of the semiclassical equations for the metric (in the weak
field approximation), keeping only terms up to second adiabatic order;
namely 
\begin{eqnarray}
&&\frac{1}{8\pi G}\left\{\Lambda{}^{(4)}g_{00}+\xi G_{00}\right\}=\langle
T_{00}(x)\rangle=\langle T_{00}(x)\rangle_{ren}+\langle T_{00}(x)\rangle^{%
\text{ad}(0)}+\langle T_{00}(x)\rangle^{\text{ad}(2)},  \label{eq00} \\
&&\frac{1}{8\pi G}\left\{G_{0i}-(\lambda-1)\partial_iK^j_j\right\}=\langle
T_{0i}(x)\rangle=\langle T_{0i}(x)\rangle_{ren}+\langle T_{0i}(x)\rangle^{%
\text{ad}(2)},  \label{eq0i}
\end{eqnarray}
where we have added and subtracted the adiabatic expansion of $\langle
T_{0\mu}(x)\rangle$ in order to separate the renormalized part $\langle
T_{0\mu}(x)\rangle_{ren}$ and the divergent contributions. The latter are to
be absorbed into a redefinition of $\Lambda$, $G$, $\lambda$ and $\xi$.
Then, we introduce Eqns. (\ref{t00ad0}) and (\ref{T0iad2}) into (\ref{eq00})
and (\ref{eq0i}), and we find that $\langle T_{00}(x)\rangle^{\text{ad}(0)}$
and $\langle T_{0\mu}(x)\rangle^{\text{ad}(2)}$ can be cancelled with the
following choice of the bare constants:

\begin{subequations}
\label{ct}
\begin{align}
\Lambda G^{-1}&=(\Lambda G^{-1})_R -\frac{\mu^{4-d}\Omega_{d-1}}{(2\pi)^{d-2}%
}I_0, \\
\xi G^{-1}&=(\xi G^{-1})_R -\frac{\mu^{4-d}\Omega_{d-1}}{6(2\pi)^{d-2}}I_1,
\\
G^{-1}&=(G^{-1})_R-\frac{\mu^{4-d}\Omega_{d-1}}{6(2\pi)^{d-2}}\left[I_2+%
\frac{2 I_3}{(d-1)(d+1)}\right], \\
G^{-1}(\lambda-1)&=(G^{-1}(\lambda-1))_R +\frac{\mu^{4-d}\Omega_{d-1}}{%
2(2\pi)^{d-2}}\frac{ I_3}{(d-1)(d+1)},
\end{align}%
where we denote the renormalized constants by a subscript $R$.

It is worth noting that from these equations we can recover the well-known
results \cite{birrell} corresponding to the usual ($z=1$) scalar field by setting $g_1=1/2$
and $g_2=g_3=0$ before taking the limit $d\to 4$. In such a case, $I_3$
vanishes and in the limit $d\to4$ we have $I_0\sim m^4(d-4)^{-1}/4$ and $%
I_1=I_2\sim m^2/(d-4)$.

All the integrals on the right hand side in Eqs. (\ref{ct}) are divergent in
the limit $d\to 4$. In the particular case of a massless field ($m=0$) these
integrals can be computed explicitly. We assume that $g_3>0$ and $%
g_2^2-4g_1g_3>0$ in order to avoid zeros of $\omega_k$. Thus, in the limit $%
d\to 4$ we have: 
\end{subequations}
\begin{subequations}
\begin{align}
i_0&\equiv \frac{\mu^{4-d}\Omega_{d-1}}{(2\pi)^{d-1}}I_0=-\frac{%
g_2(g_2^2-4g_1g_3)}{8\sqrt{2}\pi^2 g_3^{5/2}}\left[\frac{1}{d-4}-\ln(\mu
g_3^{1/4})\right]+FP ,  \notag \\
i_1&\equiv \frac{\mu^{4-d}\Omega_{d-1}}{(2\pi)^{d-1}}I_1=-\frac{%
(g_2^2-4g_1g_3)}{4\sqrt{2}\pi^2 g_3^{3/2}}\left[\frac{1}{d-4}-\ln(\mu
g_3^{1/4})\right]+FP,  \notag \\
i_2&\equiv \frac{\mu^{4-d}\Omega_{d-1}}{(2\pi)^{d-1}}I_2=-\frac{1}{\sqrt{2
g_3}\pi^2 }\left[\frac{1}{d-4}-\ln(\mu g_3^{1/4})\right]+FP ,  \notag \\
i_3&\equiv \frac{\mu^{4-d}\Omega_{d-1}}{(2\pi)^{d-1}}I_3=-\frac{2}{5 \sqrt{2
g_3}\pi^2 }\left[\frac{1}{d-4}-\ln(\mu g_3^{1/4})\right]+FP,  \notag
\end{align}%
where $FP$ denotes the $\mu$-independent finite part.

The renormalization group equations are obtained simply recalling that the
bare constants are independent of $\mu $, and are given by 
\end{subequations}
\begin{subequations}
\begin{align}
\mu \frac{d}{d\mu }\left( \Lambda G^{-1}\right) _{R}& =\frac{g_{2}}{\sqrt{2}}%
\frac{(g_{2}^{2}-4g_{1}g_{3})}{4\pi g_{3}^{5/2}}, \\
\mu \frac{d}{d\mu }\left( \xi G^{-1}\right) _{R}& =\frac{1}{\sqrt{2}}\frac{%
(g_{2}^{2}-4g_{1}g_{3})}{12\pi g_{3}^{3/2}}, \\
\mu \frac{d}{d\mu }\left( G^{-1}\right) _{R}& =\frac{3}{5\sqrt{2g_{3}}\pi },
\\
\mu \frac{d}{d\mu }\left( G^{-1}(\lambda -1)\right) _{R}& =-\frac{2}{5\sqrt{%
2g_{3}}\pi }.
\end{align}%
Note that, in contrast to what happens in the case of a usual ($z=1$) scalar
field, here we have obtained that the renormalized constants depend on $\mu $
for a massless field. Note also that in these equations, as $g_{3}>0$ and $%
g_{2}^{2}-4g_{1}g_{3}>0$, the right hand side have a determined sign, except
in the first equation which depends on the sign of $g_{2}$.

Assuming that the order of magnitude of the constants $g_i$ are
determined by a single mass scale $M_C$, i.e. $g_1\sim g_2 M_C^2\sim g_3 M_C^4\sim \mathcal{O}(1)$,
these equations give us information on the running of the
renormalized constants in the UV \cite{ShapiroBook}, at scales larger than $M_C$.
With the renormalization scheme we are
considering (minimal subtraction), it is not possible to analyze the IR
behaviour of the coupling constants. As pointed out in \cite{CollinsRen,Russo,Shapiro}, this would require a mass dependent
renormalization scheme. We hope to address
this problem in a forthcoming publication.

\begin{acknowledgments}
This work was supported by Universidad de Buenos Aires UBA, CONICET\ and
ANPCyT. The authors thank Diego Blas, Oriol Pujol\`{a}s and Jorge Russo for
interesting discussions.
\end{acknowledgments}

\end{subequations}


\begin{thebibliography}{99}
\bibitem{Horava} P. Ho\v{r}ava, Phys. Rev. \textbf{D79} (2009) 084008,
[arXiv:0901.3775].



\bibitem{Blas} D. Blas, O. Pujol\`{a}s and S. Sibiryakov, JHEP \textbf{0910}
(2009) 029, [arXiv:0906.3046].

\bibitem{Charmousis} C. Charmousis, G. Niz, A. Padilla and P. Saffin, JHEP 
\textbf{0908} (2009) 070, [arXiv:0905.2579].

\bibitem{Li} Miao Li and Yi Pang, JHEP \textbf{0908} (2009) 015,
[arXiv:0905.2751].

\bibitem{Henneaux} M. Henneaux, A. Kleinschmidt and G. Lucena G\'{o}mez,
Phys. Rev. \textbf{D81} (2010) 064002, [arXiv:0912.0399].

\bibitem{Blas2} D. Blas, O. Pujol\`{a}s and S. Sibiryakov, [arXiv:0909.3525].

\bibitem{Blas3} D. Blas, O. Pujol\`{a}s, S. Sibiryakov, Phys. Lett. \textbf{%
B688} (2010) 350, [arXiv:0912.0550].

\bibitem{DiegoDiana} D. L\'{o}pez Nacir and F. D. Mazzitelli, Phys. Lett. 
\textbf{B672} (2009) 294, [arXiv:0810.2922].




\bibitem{Jacobson} T. Jacobson, [arXiv:1001.4823].

\bibitem{Germani} C. Germani, A. Kehagias and K. Sfetsos, [arXiv:0906.1201].

\bibitem{Brandenberger} Xian Gao, Yi Wang, R. Brandenberger and A. Riotto,
Phys. Rev. \textbf{D81} (2010) 083508, [arXiv:0905.3821].

\bibitem{Anselmi} D. Anselmi and M. Halat, Phys. Rev. {\bf D76} (2007) 125011.

\bibitem{Visser} M. Visser, [arXiv:0912.4757].

\bibitem{CollinsRen} J. Collins, \textit{Renormalization}, Cambridge
University Press, Cambridge, 1984.

\bibitem{birrell}N. D. Birrell and P. C. W. Davies, {\it Quantum Fields in
Curved Space} (Cambridge University Press, Cambridge, 1982).

\bibitem{ShapiroBook} I.L. Buchbinder, S.D. Odintsov and I.L. Shapiro, \textit{Effective action in
quantum gravity}, Institute of Physics Publishing, Bristol and Philadelphia, 1992.

\bibitem{Russo} R. Iengo, J. Russo and M. Serone, JHEP \textbf{0911} (2009)
020, [arXiv:0906.3477].

\bibitem{Shapiro} E. Gorbar and I. Shapiro, JHEP \textbf{0302} (2003) 021.

\end{thebibliography}
\end{document}